\documentstyle[11pt]{article}

\renewcommand{\theequation}{\thesection.\arabic{equation}}

\newlength{\extraspace}
\newlength{\extraspaces}
\newcounter{dummy}

\newcommand{\baa}{
\addtocounter{equation}{1}
\setcounter{dummy}{\value{equation}}
\setcounter{equation}{0}
\renewcommand{\theequation}{\thesection.\arabic{dummy}\alph{equation}}
\begin{eqnarray}
\addtolength{\abovedisplayskip}{\extraspaces}
\addtolength{\belowdisplayskip}{\extraspaces}
\addtolength{\abovedisplayshortskip}{\extraspace}
\addtolength{\belowdisplayshortskip}{\extraspace}}

\newcommand{\eaa}{
\end{eqnarray}
\setcounter{equation}{\value{dummy}}
\renewcommand{\theequation}{\thesection.\arabic{equation}}}

\newcommand{\be}{\begin{equation}
\addtolength{\abovedisplayskip}{\extraspaces}
\addtolength{\belowdisplayskip}{\extraspaces}
\addtolength{\abovedisplayshortskip}{\extraspace}
\addtolength{\belowdisplayshortskip}{\extraspace}}
\newcommand{\ee}{\end{equation}}

\newcommand{\ba}{\begin{eqnarray}
\addtolength{\abovedisplayskip}{\extraspaces}
\addtolength{\belowdisplayskip}{\extraspaces}
\addtolength{\abovedisplayshortskip}{\extraspace}
\addtolength{\belowdisplayshortskip}{\extraspace}}
\newcommand{\ea}{\end{eqnarray}}

\newcommand{\bd}{\begin{displaymath}
\addtolength{\abovedisplayskip}{\extraspaces}
\addtolength{\belowdisplayskip}{\extraspaces}
\addtolength{\abovedisplayshortskip}{\extraspace}
\addtolength{\belowdisplayshortskip}{\extraspace}}
\newcommand{\ed}{\end{displaymath}}

\newcommand{\ban}{\begin{eqnarray*}
\addtolength{\abovedisplayskip}{\extraspaces}
\addtolength{\belowdisplayskip}{\extraspaces}
\addtolength{\abovedisplayshortskip}{\extraspace}
\addtolength{\belowdisplayshortskip}{\extraspace}}
\newcommand{\ean}{\end{eqnarray*}}

\newcommand{\newsection}[1]{
\vspace{10mm}
\pagebreak[3]
\addtocounter{section}{1}
\setcounter{equation}{0}
\setcounter{subsection}{0}
\setcounter{footnote}{0}
\begin{center}
{\Large \thesection. #1}
\end{center}
\nopagebreak
\medskip
\nopagebreak}

\newcommand{\newapp}{
\vspace{5mm}
\pagebreak[3]
\addtocounter{section}{1}
\setcounter{equation}{0}
\setcounter{subsection}{0}
\setcounter{footnote}{0}
\begin{flushleft}
{\bf Appendix \thesection}
\end{flushleft}
\nopagebreak
\medskip
\nopagebreak}

\newcommand{\startappendix}{
\renewcommand{\thesection}{\Alph{section}}
\setcounter{section}{0}}

\newcommand{\newsubsection}[1]{
\vspace{8mm}
\pagebreak[3]

\addtocounter{subsection}{1}
\noindent{ \sc \thesubsection. #1}
\nopagebreak
\vspace{2mm}
\nopagebreak}

\newcommand{\nonu}{\nonumber \\[.5mm]}
\newcommand{\U}[5]{{#5}^{a_{#1}\ldots a_{#2}}_{b_{#3}\ldots b_{#4}}}

\newcommand{\hu}{\widetilde{U}}

\newcommand{\deel}[2]{{\textstyle{#1 \over #2}}}
\newcommand{\hf}{{\textstyle{1\over 2}}}

\newcommand{\ie}{{\it i.e.}}

\newcommand{\re}{\mbox{I}\!\mbox{R}}

\def\inbar{\,\vrule height1.5ex width.4pt depth0pt}
\font\rms=cmr12 at 12pt
\def\ce{\relax\ifmmode\mathchoice
{\hbox{$\inbar\kern-.3em{\rm C}$}}
{\hbox{$\inbar\kern-.3em{\rm C}$}}
{\lower.9pt\hbox{\rms $\inbar\kern-.3em{\rm C}$}}
{\lower1.2pt\hbox{\rms $\inbar\kern-.3em{\rm C}$}}
\else{$\inbar\kern-.3em{\rm C}$}\fi}
\font\cmss=cmss12 \font\cmsss=cmss12 at 12pt
\def\ze{\relax\ifmmode\mathchoice
{\hbox{\cmss Z\kern-.4em Z}}{\hbox{\cmss Z\kern-.4em Z}}
{\lower.9pt\hbox{\cmsss Z\kern-.4em Z}}
{\lower1.2pt\hbox{\cmsss Z\kern-.4em Z}}\else{\cmss Z\kern-.4em Z}\fi}

\newcommand{\dif}{\partial}
\newcommand{\dbar}{\bar{\dif}}

\newcommand{\bgamma}{\bar{\gamma}}

\newcommand{\bzeta}{\bar{\zeta}}
\newcommand{\etab}{\bar{\eta}}
\newcommand{\bbeta}{\bar{\beta}}

\newcommand{\lamb}{\bar{\lambda}}

\newcommand{\beps}{\bar{\epsilon}}
\newcommand{\bz}{\bar{z}}

\newcommand{\bw}{\bar{w}}

\newcommand{\gam}{\Gamma}
\renewcommand{\ll}{\Lambda}

\newcommand{\tr}[1]{\mbox{Tr}#1}
\newcommand{\del}{\delta}

\newcommand{\om}{\omega}
\newcommand{\bom}{\bar{\omega}}
\newcommand{\bc}{\bar{c}}
\newcommand{\bb}{\bar{b}}
\newcommand{\vr}{\delta_{\epsilon,\bar{\epsilon}}}
\newcommand{\vrr}{\delta_{brst}}

\newcommand{\mub}{\bar{\mu}}
\newcommand{\nub}{\bar{\nu}}

\newcommand{\ac}[1]{\deel{1}{2\pi}\int d^2z \, \tr{#1}}

\newcommand{\aksie}[1]{\deel{1}{\pi}\int d^2z \, \left[ {#1}
\right]}
\newcommand{\aksies}[1]{\deel{k}{\pi}\int d^2z \, \left[ {#1}
\right]}

\newcommand{\var}[1]{{\del \over \del {#1}}}
\newcommand{\vars}[2]{{\del {#1} \over \del {#2}}}

\newcommand{\mat}[9]{\left( \begin{array}{ccc}
				#1 & #2 & #3 \\
				#4 & #5 & #6 \\
				#7 & #8 & #9
                             \end{array} \right) }

\newcommand{\mats}[4]{\left( \begin{array}{cc}
				#1 & #2 \\
				#3 & #4
                             \end{array} \right) }

\newcommand{\np}[1]{Nucl. Phys. {\bf B#1}}
\newcommand{\cmp}[1]{Comm. Math. Phys. {\bf #1}}

\newcommand{\plb}[1]{Phys. Lett. {\bf B#1}}

\newcommand{\eps}{\epsilon}

\begin{document}
\addtolength{\baselineskip}{.7mm}

\thispagestyle{empty}
\begin{flushright}
{\sc THU}-92/34\\
11/92
\end{flushright}
\vspace{1.5cm}
\setcounter{footnote}{2}
\begin{center}
{\LARGE\sc{KPZ Analysis for $W_3$ Gravity}}\\[2cm]

\sc{Jan de Boer\footnote{e-mail: deboer@ruunts.fys.ruu.nl}
and Jacob Goeree\footnote{e-mail: goeree@ruunts.fys.ruu.nl}}
\\[8mm]
{\it Institute for Theoretical Physics\\[2mm]
University of Utrecht\\[2mm]
Princetonplein 5\\[2mm]
P.O. Box 80.006\\[2mm]
3508 TA Utrecht}\\[1.5cm]

{\large\sc Abstract}\\[1cm]
\end{center}

\noindent Starting from the covariant action for $W_3$ gravity
constructed in \cite{jj}, we discuss the BRST quantization of
$W_3$ gravity. Taking the chiral gauge the
BRST charge has a natural interpretation in terms of the quantum
Drinfeld--Sokolov reduction for $Sl(3,\re)$. Nilpotency of this
charge leads to the KPZ formula for $W_3$. In the conformal
gauge, where the covariant action reduces to a Toda action, the
BRST charge is equivalent to the one recently constructed in
\cite{cern}.
\vfill

\newpage
\newsection{Introduction}

Consider some matter fields coupled to gravity in a
diffeomorphism and Weyl invariant way.
Integrating out the matter from such a theory one
obtains a gravitational induced action $\Gamma[g_{ab}]$.
If the theory has no anomalies, $\Gamma[g_{ab}]$
reduces to an action on moduli space, since on the
classical level the number of degrees of freedom of the metric
equals the number of invariances. However, as is well known,
the procedure of integrating out the matter cannot be
done in both a Weyl and diffeomorphism invariant way,
leading to a non-trivial $g$ dependence of $\Gamma[g_{ab}]$.
Using a diffeomorphism invariant regulator to handle
the matter integration one obtains the following unique
expression for the induced action
\be \label{polyak}
\Gamma[g_{ab}]=\frac{d}{96\pi} \int R \, \frac{1}{\Box} R,
\ee
a result first obtained by Polyakov \cite{pol1}. Note
that $d$, the central charge of the matter system, is the
only remnant of the matter system we used to define the
induced action.
Parametrizing the metric by $ds^2=e^{-2\phi}|dz+\mu d\bz|^2$,
the induced action $\Gamma[g_{ab}]$ decomposes into a sum of
three terms \cite{herman}:
\be \label{drietermen}
\Gamma[g_{ab}]=\Gamma[\mu]+\Delta
S[\mu,\mub,\phi]+\Gamma[\mub],
\ee
where $\Gamma[\mu]$ and $\Gamma[\mub]$ are non-local actions,
known as the {\em chiral}\, actions, while $\Delta S$ is a {\em
local}\, counterterm, which is such that the sum of the three
terms is invariant under diffeomorphisms. Using the operator
product expansion of the stress-energy tensor $T_{mat}$ with
itself, the chiral action $\Gamma[\mu]$ can be computed from
\be \label{defchir}
\exp(-\Gamma[\mu])=\left< \exp\deel{1}{\pi}\int \mu T_{mat}
\right>_{OPE}.
\ee
A similar definition holds for $\Gamma[\mub]$.
In \cite{pol3,herman} it was shown that in the semi-classical limit
$d\rightarrow \infty$ the chiral action $\Gamma[\mu]$ is
related to the WZW action based on $Sl(2,\re)$
\be \label{introrelatie}
\exp(-\gam[\mu])=\int Dh\,\delta(k\mu-(J_++2zJ_0-z^2J_-))
\,\exp(-kS^+_{wzw}(h)),
\ee
with $J=kh^{-1}\dbar h$ and $k\sim d/6$.

In \cite{jj1,jj2} we generalized the above results to $W_N$
gravity. Starting with the chiral action
\be \label{wnchir}
\Gamma[\mu_i]=\left< \exp\deel{1}{\pi}\int \sum_{i=2}^N
\mu_i W_{mat}^i\right>_{OPE},
\ee
where the $\mu_i$ are the $W_N$ generalizations of the Beltrami
differential $\mu_2$, we
constructed the local counterterm $\Delta S$ such that
\be \label{pr}
S_{cov}=\gam[\mu_i]+\Delta S[\mu_i,\mub_i,G]+\gam[\mub_i],
\ee
is invariant under left and right $W_N$ transformations. Here
$G$ is a $Sl(N,\re)$ valued field needed to make the action
invariant. Taking a Gauss decomposition for $G$ it turns out
that some of the components of $G$ are auxiliary fields that can
be integrated out \cite{jj1,jj2}. For instance, for $Sl(2,\re)$
only the Cartan subgroup labels a true degree of freedom, which
is the Liouville field $\phi$.
Furthermore, it was shown that in the limit $d\rightarrow
\infty$ the chiral actions are related to the WZW action based on
$Sl(N,\re)$ in a way similar as (\ref{introrelatie}), and that
the covariant action $S_{cov}$ is nothing but a Legendre
transform of the WZW action \cite{jj2}.

In this paper we will discuss the quantization of this theory
for the case of $Sl(3,\re)$, \ie\ $W_3$ gravity.
Previously, there have been attempts to quantize $W_3$ gravity from
the chiral point of view \cite{vannie} (see also
\cite{wreview1,wreview2,wreview3}). For example Schoutens
{\em et. al.}\, \cite{vannie} computed the one-loop contributions
to the effective action for $W_3$ gravity, using only the chiral
action $\Gamma[\mu,\nu]$ (where we denoted the
$W_3$ analogue of the Beltrami differential by $\nu$).
Here we take a different point of view and discuss the
{\em covariant}\, quantization of $W_3$ gravity. Starting from
the covariant action $S_{cov}$ specialized to $Sl(3,\re)$,
we will impose the chiral gauge using the BRST formalism. This
procedure gives us a {\em classical}\, BRST charge, whose
nilpotency can be checked using the Poisson brackets of the
fields contained in the BRST charge.
Quantization of the theory is subsequently done in the following
way: the Poisson brackets are replaced by OPE's, the classical BRST
charge is replaced by a normal-ordered one, and
we allow for multiplicative renormalizations of the fields
contained in the BRST charge. Nilpotency of this quantum BRST charge
will lead to the KPZ formula for $W_3$.

We will illustrate our method in section~2
for the case of ordinary gravity.
We show that BRST quantization of ordinary gravity in the chiral
gauge $\mub=\phi=0$ leads to a BRST charge of the form
\be \label{introvorm}
Q=Q_{ds}+Q_{FF},
\ee
with
\ba \label{introwith}
Q_{ds} &=& \oint c(J_--k),\nonu
Q_{FF} &=& \oint :\bc\left( T_{mat}+T_{FF}+\hf T_{gh} \right):,
\ea
with $\oint=\deel{1}{2\pi i}\oint d\bz$.
Here $Q_{ds}$ follows from the gauge-fixing of $\phi$;
it is the BRST charge that is naturally associated
with the Drinfeld--Sokolov reduction for $Sl(2,\re)$. $Q_{FF}$
follows from the gauge-fixing of $\mub$, and $T_{FF}$ appearing
in $Q_{FF}$ is the generator of the cohomology of $Q_{ds}$.
Nilpotency of $Q$, and in fact
\be \label{introfact}
Q_{ds}^2=Q_{ds}Q_{FF}+Q_{FF}Q_{ds}=Q_{FF}^2=0,
\ee
holds iff
\be \label{introkpz}
d=13+6\left( (k+2)+\frac{1}{(k+2)} \right),
\ee
which is the well-known KPZ result for ordinary gravity \cite{kpz}.

In section~3 these results are generalized to the case of $W_3$ gravity
in the following way. Starting with the covariant action and
subsequently imposing the chiral gauge we will derive the BRST
charge for $W_3$. It again takes the form $Q=Q_{ds}+Q_{FF}$,
with $Q_{ds}$ the BRST charge associated with the
Drinfeld--Sokolov reduction for $Sl(3,\re)$, and
\be \label{introw3ff}
Q_{FF}=\oint :\bc_1 \left( T_{mat}+T_{FF}+\hf T_{gh} \right)+
\bc_2 \left( \pm i\widetilde{W}_{mat}+\widetilde{W}_{FF}+
\hf \widetilde{W}_{gh} \right):,
\ee
where $T_{FF},\widetilde{W}_{FF}$ are the generators of the $Q_{ds}$
cohomology. Nilpotency of $Q$ now holds iff
\be \label{introkpzw3}
d=50+24\left( (k+3)+\frac{1}{(k+3)} \right),
\ee
which is the KPZ result for $W_3$ gravity, previously
conjectured in \cite{matsuo,beroog,kj1}.

It should be stressed that the existence of a BRST charge for
$W_3$ matter coupled to $W_3$ gravity is somewhat surprising, since one
of the basic requirements for the construction of a BRST charge
is that the generators contained in the BRST charge form a closed
algebra. (The structure coefficients of the algebra may be field
dependent. As long as the algebra closes one can in general
construct the BRST charge.) But for $W_3$ matter coupled to
$W_3$ gravity this is {\em not}\, the case. Although $W_{mat}$
and $W_{FF}$ form a closed $W_3$ algebra seperately,
a linear combination of them
does not, due to the non-linearity of the
$W_3$ algebra. Nevertheless, there exists a BRST charge for this
coupled system. Its existence can be justified
as follows: since there is a covariant action describing $W_3$ matter
coupled to $W_3$ gravity, we know that the BRST quantization of
this action should lead to the BRST charge for $W_3$ matter
coupled to $W_3$ gravity.

Recently, there have been other constructions of the BRST
charge \cite{cern,cern2} for $W_3$ matter coupled to $W_3$ Toda
theory. In section~4 we will discuss the BRST quantization of
the covariant action in the conformal gauge, where it reduces to
a Toda theory, and show that in this gauge our BRST charge
equals that of \cite{cern,cern2}.

\newpage

\newsection{Covariant Quantization of Ordinary Gravity}

In this section we will perform the BRST quantization of
ordinary gravity. We start with a covariant formulation
of 2D gravity and subsequently impose the chiral gauge
using the BRST formalism. We argue that the corresponding
BRST charge can be understood from the point of view of
(quantum) Drinfeld--Sokolov reductions \cite{DS,FF,beroog}.
Using this connection we will in the next section discuss
the quantization of $W_3$ gravity. Previous studies concerning
the BRST quantization of 2D gravity have appeared in
\cite{horva,itoh,kura}.

The covariant action for ordinary gravity,
given by the general formula (\ref{pr}) specialized to the case
of $Sl(2,\re)$, reads:
\be \label{global}
S_{cov}=\gam[\mu]+\Delta S[\mu,\mub,G]+\gam[\mub].
\ee
If we take the following Gauss decomposition for $G$
\be \label{chauss}
G=\mats{1}{0}{\om}{1}\mats{e^{\phi}}{0}{0}{e^{-\phi}}
\mats{1}{-\bom}{0}{1},
\ee
the local counterterm $\Delta S$ becomes \cite{herman,jj1}
\be \label{w2aksie}
\Delta S=\deel{k}{\pi}\int d^2z\Bigl[
\dif \phi \dbar \phi + \omega(2\dbar \phi +\dif \mu)
+\bar{\omega} ( 2\dif \phi + \dbar \bar{\mu})
+\mu \omega^2 + \bar{\mu}\bar{\omega}^2 +
2\omega\bar{\omega}-(1-\mu\bar{\mu})e^{-2\phi} \Bigr].
\ee
As we already mentioned in the Introduction, the relation between
$\gam[\mu]$ and the $Sl(2,\re)$ WZW model reads \cite{pol3,herman,jj2}:
\be \label{relatie}
\exp(-\gam[\mu])=\int Dh\,\delta(k\mu-(J_++2zJ_0-z^2J_-))
\,\exp(-kS^+_{wzw}(h)),
\ee
where we decomposed the $Sl(2,\re)$ current $J=kh^{-1}\dbar h$ as
\be \label{tweedecom}
J=J_aT^a=\mats{J_0}{J_+}{J_-}{-J_0}.
\ee
The definitions of the WZW actions $S^{\pm}_{wzw}$
can be found in Appendix~A.
Note that the equations of motion
$\dif J_a=0$ for the currents $J_a$, that
follow from the WZW action,
lead to the correct equation of motion for $\mu$, \ie\ $\dif^3
\mu=0$. Classically, the $J_a$ satisfy the Poisson
brackets\footnote{Our conventions are: $\eta^{ab}=\mbox{Tr}(T^aT^b)$,
$\eta_{ab}$ is the inverse of $\eta^{ab}$, $f^{ab}_cT^c=[T^a,T^b]$,
and indices are raised using $\eta^{ab}$.}
\be \label{pbs}
\{ J_a(\bz),J_b(\bw)\}=k\eta_{ab}\del'(\bz-\bw)+f^c_{ab}J_c(\bw)
\del(\bz-\bw),
\ee
where the prime means derivation w.r.t. $\bw$. These Poisson brackets
should be replaced by the Operator Product Algebra (OPA)
\be \label{OPA1}
J_a(\bz)J_b(\bw)=\frac{k\eta_{ab}}{(\bz-\bw)^2}+
\frac{f_{ab}^cJ_c(\bw)}{(\bz-\bw)},
\ee
when quantizing the theory.

In (\ref{w2aksie})
$\om,\bom$ are auxiliary fields that can be integrated out.
If one does so \cite{herman,jj1},
one recovers Polyakov's result for the (induced)
gravitational action \cite{pol1}:
$S_{cov}\sim \int R\,\frac{1}{\Box}R$,
with $\Box$ the covariant Laplacian and
$R$ the scalar curvature corresponding to the metric defined
by $ds^2=e^{-2\phi}|dz+\mu d\bar{z}|^2$.
We prefer not to do this, because in the case of $W$ gravity
the auxiliary fields appear in general in more than second
order and can thus not be integrated out \cite{jj1}.
The covariant action (\ref{global}) is invariant under general
two-dimensional diffeomorphisms, whose action on the fields is
given by \cite{jj1,jj2}:
\ba \label{trafos}
\vr \mu &=& \dbar \eps+\eps \dif \mu -\mu \dif \eps, \nonu
\vr \mub &=& \dif \beps+\beps \dbar \mub -\mub \dbar \beps, \nonu
\vr \phi &=& -\hf\dif\eps -\hf\dbar\beps -\om\eps -\bom \beps, \nonu
\vr \om &=& \hf \dif^2 \eps+\dif(\eps \om)-\mub \eps
e^{-2\phi}+\beps e^{-2\phi}, \nonu
\vr \bom &=& \hf \dbar^2 \beps+\dbar(\beps\bom)-\mu \beps
e^{-2\phi}+\eps e^{-2\phi}.
\ea

\newsubsection{BRST Quantization}

We now study this theory in the chiral gauge
$\mub=\phi=0$. This can be done most easily using the BRST
formalism. We replace the above transformation rules
of the fields by BRST transformations, \ie\ we replace
$\eps,\beps$ in (\ref{trafos}) by ghost fields $c,\bc$,
supplemented with the rule $\vrr c=c\dif c$ and $\vrr \bc
=\bc \dbar \bc$. In order to impose the gauge we introduce
two more anti-ghost fields $b,\bb$, whose BRST transformation
rules are $\vrr b=B$ and $\vrr \bb=\bar{B}$. The gauge-fixed
action reads
\be \label{gfaksie}
S_{gf} = S_{cov}-\vrr \left( \aksie{\bb \mub-2 b \phi} \right).
\ee
Obviously, this action is invariant
under the above BRST transformations, since the classical action
was BRST invariant from the start and for the gauge fixing term
we have\footnote{Note that
it is not true that $\vrr^2$ vanishes off-shell on all the
fields. In particular $\vrr^2\om$ and $\vrr^2\bom$ are
proportional to the equations of motion of $\om,\bom$.
Since these fields do not appear in the gauge fixing term, this
does not cause any trouble.}
$\vrr^2(\mbox{gauge fixing term})=0$.
Integrating out the Nakanishi--Lautrup fields $B,\bar{B}$
imposes the gauge fixing condition $\mub=\phi=0$, and changes
the BRST transformation of $b,\bb$ into
\ba \label{modbrule}
\vrr b &=& -\frac{\pi}{2} \vars{S_{gf}}{\phi} \,\,=\,\,
k(\dbar \om +\dif \bom -1), \nonu
\vrr \bb &=& \pi \vars{S_{gf}}{\mub} \,\,=\,\,
T_{mat}+T_{\Delta}+T_{gh},
\ea
where we defined
\ba \label{defs}
T_{mat} &=& \pi\vars{\gam[\mub]}{\mub}, \nonu
T_{\Delta} &=& \pi\vars{\Delta S}{\mub}=k(\bom^2-\dbar\bom+
\mu),\\[.8mm]
T_{gh} &=&-\bb\dbar\bc-\dbar(\bb\bc). \nonumber
\ea
The gauge-fixed action becomes:
\be \label{gfaksie2}
S_{gf}=\gam[\mu]+\aksies{\om\dif \mu+\mu\om^2+2\om\bom}+
\aksie{\bb\dif\bc+b(\dif c+\dbar \bc+2c\om+2\bc\bom)}.
\ee

At this point we would like to integrate out the auxiliary fields
$\om,\bom$. Notice that the term quadratic in $\bom$ has
disappeared. This is something which generalizes to $W_3$ gravity:
after taking the `chiral' gauge, whose definition for $W_3$
gravity is given in
the next section, the troublesome terms of higher than quadratic
order in the auxiliary fields disappear making it possible to
integrate them out. In this case the equations of motion for
$\om,\bom$ are: $\om=-\frac{1}{k}b\bc$ and
$\bom=-\hf\dif\mu+\frac{1}{k}(\mu b\bc-bc)$.
Substituting these equations of motion leaves us with the
following action:
\be \label{gfaksie3}
S_{gf}=\gam[\mu]+\aksie{\bb\dif\bc+b\dif c+b\dbar\bc-b\bc \dif \mu},
\ee
which is of course still BRST invariant. The ghost
action, which is off-diagonal in this form, can be diagonalized
by making a ghost field redefinition
\ba \label{newshoes}
\gamma=c-\mu \bc+z\dbar \bc,&\hspace{7mm}&\bgamma=\bc, \nonu
\bbeta=\bb+\mu b+z\dbar b,&\hspace{7mm}&\beta=b.
\ea
In terms of these new variables the final action reads
\be \label{finalaksie}
S_{final}=\gam[\mu]+\aksie{\bbeta\dif\bgamma+\beta\dif\gamma}.
\ee
Note that the new ghost fields have the same `diagonal'
Poisson brackets
\be \label{diagopbs}
\{ \bbeta(\bz),\bgamma(\bw)\}=\del(\bz-\bw),\hspace{1cm}
\{ \beta(\bz),\gamma(\bw)\}=\del(\bz-\bw)
\ee
as the old ones. Also these Poisson brackets should be replaced
by OPE's when quantizing the theory.

If we use the correspondence (\ref{relatie}) between $\mu$ and the
currents $J_a$ of the $Sl(2,\re)$ current algebra, we arrive
at the following {\em on-shell}\, BRST transformation
rules\footnote{Note that on-shell fields
are {\em anti-holomorphic}, \ie\ $\dif(\mbox{fields})=0$.}
\cite{kura}:
\ba \label{finalbrst}
\vrr J_- &=& (\dbar J_-) \bgamma, \nonu
\vrr J_0 &=& -J_-\gamma+\dbar(J_0\bgamma)-\deel{k}{2}\dbar^2\bgamma,
\nonu
\vrr J_+ &=& 2J_0\gamma+k\dbar\gamma
+(\dbar J_+)\bgamma+2J_+\dbar \bgamma,
\nonu
\vrr \bgamma &=& \bgamma\dbar\bgamma, \nonu
\vrr \gamma  &=& -\dbar(\gamma\bgamma), \\[.5mm]
\vrr \beta   &=& J_--k-(\dbar\beta)\bgamma, \nonu
\vrr \bbeta  &=& T_{mat}+T_{sug}+T_{gh}, \nonumber
\ea
where $T_{sug}$ is an improved version of the classical
Sugawara stress-energy tensor
\be \label{sugimpr}
T_{sug}=\frac{1}{2k}\eta^{ab}J_aJ_b+\dbar J_0,
\ee
and $T_{gh}=T_{gh}^{\bbeta,\bgamma}+T_{gh}^{\beta,\gamma}=
-\bbeta\dbar\bgamma-\dbar(\bbeta\bgamma)+(\dbar
\beta)\gamma$.

It is straightforward to show that these BRST transformations
are generated by the BRST charge
\be \label{brst1}
Q=\oint \gamma(J_--k)+\bgamma\left( T_{mat}+T_{sug}+\hf
T_{gh}^{\bbeta,\bgamma}+T_{gh}^{\beta,\gamma}\right).
\ee
Gauge independence of the theory is guaranteed by nilpotency
of the BRST charge $Q$. Classically, using the
Poisson brackets (\ref{pbs}) and (\ref{diagopbs}) to evaluate $Q^2$,
this gives the following relation
between the matter central charge $d$ and the level $k$ of the
$Sl(2,\re)$ current algebra: $d-6k=0$. At the quantum level
things are a bit more complicated: first, we
should replace the BRST operator by a normal ordered version,
and second, we allow for possible renormalizations of the
stress-energy tensor $T_{sug}$ appearing in the BRST charge:
\be \label{brst2}
Q_{quant}=\oint :\gamma(J_--k)+\bgamma\left(
T_{mat}+T_{sug}+\hf T_{gh}^{\bbeta,\bgamma}+
T_{gh}^{\beta,\gamma}\right):,
\ee
with
\be \label{quantsug}
T_{sug}=\hf N_T\, \eta^{ab}:J_aJ_b:+x \dbar J_0.
\ee
Here $N_T,x$ should follow from the requirement of nilpotency of
the quantum BRST charge. Using the operator product
expansions for the ghosts and currents
one easily verifies that
the BRST charge is only nilpotent when the following
criteria are satisfied
\be \label{crit}
x=1,\hspace{1cm}N_T=\frac{1}{k+2},\hspace{1cm}d+\frac{3k}{k+2}-6k-28=0.
\ee
Thus the requirement of diffeomorphism invariance at the quantum
level leads to a renormalization of the level $k$ of the $Sl(2,\re)$
current algebra from $k=d/6$ to the one following from
(\ref{crit}), a result first obtained by KPZ \cite{kpz}.
Note that the relation between $d$ and $k$ can also be
written as:
\be \label{rel2}
d=13+6\left( (k+2)+\frac{1}{(k+2)} \right).
\ee

\newsubsection{Relation with Quantum Drinfeld--Sokolov Reductions}

In this section we want to reinterprete the previous results
in terms of the quantum Drinfeld--Sokolov reduction of
$Sl(2,\re)$. Recall that (quantum) Drinfeld--Sokolov
reductions provide a very efficient way to
obtain (quantum) $W$ algebras \cite{DS,FF,beroog,BS}.
Let us illustrate this for the $Sl(2,\re)$ case.
Classically the approach boils down to the
following: one starts with the dual space $M$
of the level-$k$ extended loop algebra of $Sl(2,\re)$, on which
we have the Poisson structure (\ref{pbs}).
Subsequently, one defines the {\em reduced phase space}\,
$M_{red}$ by imposing the (first-class) constraint $J_-=\xi$,
with $\xi$ some constant, on the
$Sl(2,\re)$ current $J$. We will take $\xi=k$ to be in agreement
with the previous subsection, although one could leave $\xi$
a free parameter. As usual, such a first-class constraint
generates gauge invariances on the reduced phase space.
The {\em physical phase space}\, $M_{phys}$ is the set
of gauge invariant quantities built on elements of
the constrained phase space $M_{red}$, \ie\
$M_{phys}=M_{red}/{\cal G}$, where ${\cal G}$ is the symmetry
group generated by the first class constraint, acting on
$M_{red}$ via gauge transformations.
In the case of $Sl(2,\re)$ we have ${\cal G}=U^+$, \ie\ the
group of upper triangular matrices, which can be used to bring the
elements of $M_{red}$ in the form:
\be \label{form}
\mats{0}{T}{k}{0},
\ee
where $T$ is related to the original currents by:
$T=\frac{1}{k}J_0^2+J_++\dbar J_0$ (one easily checks that this $T$ is
indeed invariant under ${\cal G}$ gauge transformations).
The Poisson bracket on $M$ induces a Poisson bracket on $M_{phys}$.
Using the Poisson brackets (\ref{pbs}) an easy computation shows
that in this case $T$ is the generator of the Virasoro algebra
with central charge $-6k$
\be \label{vira}
\{ T(\bz),T(\bw)\}=-\hf
k\del'''(\bz-\bw)+2T(\bw)\del'(\bz-\bw)+T'(\bw)\del(\bz-\bw).
\ee

In the quantum analogue of this construction the constraint
$J_-=k$ is implemented using the BRST procedure. For this
one introduces the (anti)ghost $\beta,\gamma$ and the BRST charge
\be \label{ds}
Q_{ds}=\oint \gamma(J_--k).
\ee
Physical operators ${\cal O}_i$ are now characterized as
follows: $Q_{ds}({\cal O}_i)=0$, and ${\cal O}_i$
and ${\cal O}_j$ are equivalent if their difference is the
$Q_{ds}$ of something. ($Q({\cal O})$ is computed by
performing the contour integration $\deel{1}{2\pi i}
\oint_{\cal C}d\bz$ of the OPE of $\gamma(\bz)(J_-(\bz)-k)$ with
${\cal O}(\bw)$. Here ${\cal C}$ is a contour around $\bw$.)
So we are naturally interested in the cohomology of $Q_{ds}$
\be \label{coho}
H_{Q_{ds}}=\frac{\mbox{Ker}\,Q_{ds}}{\mbox{Im}\,Q_{ds}}.
\ee

Turning back to the quantum BRST charge (\ref{brst2}) we derived
in the previous subsection, we see that part of this charge is
precisely $Q_{ds}$. In fact, if we split the BRST charge
in (\ref{brst2}) as $Q=Q_{ds}+Q_{sug}$, we have a stronger
result than nilpotency of the total BRST charge $Q$, namely:
\be \label{stronger}
Q_{ds}^2=Q_{ds}Q_{sug}+Q_{sug}Q_{ds}=Q_{sug}^2=0.
\ee
Here nilpotency of $Q_{ds}$ is obvious, $Q_{ds}$ and $Q_{sug}$
anti-commute due to the fact that
\be \label{dueness}
Q_{ds}(T_{sug}+T_{gh}^{\beta,\gamma})=0,
\ee
($Q_{ds}$ obviously vanishes on $T_{mat}$ and
$T_{gh}^{\bbeta,\bgamma}$) and $Q_{sug}$ is nilpotent iff
$T_{ds}=T_{sug}+T_{gh}^{\beta,\gamma}$ generates a Virasoro algebra
with central charge $26-d$. This last requirement leads of course to
the relation (\ref{rel2}) between the level $k$ and the matter
central charge $d$. Because $T_{ds}$ generates a Virasoro
algebra with non-vanishing central charge it cannot be $Q_{ds}$
exact, which implies that $T_{ds}$ is a generator of
$H_{Q_{ds}}$. In fact, it was proven by Feigin and
Frenkel \cite{FF} that $T_{ds}$ generates the {\em whole}\,
$Q_{ds}$ cohomology.

The above makes it clear that the (quantum)
Drinfeld--Sokolov reduction for the case of $Sl(2,\re)$ gives
rise to a (quantum) Virasoro algebra.
Furthermore, we see that quantization of 2D gravity
in the chiral gauge actually resolves the cohomological
problem posed in the Drinfeld--Sokolov approach; by
quantizing the covariant action (\ref{global}) we found the
explicit form for the generator of $H_{Q_{ds}}$.
Conversely, we could use the Drinfeld--Sokolov approach as
follows: find the generator of the cohomology of $Q_{ds}$ and use
this to construct $Q_{sug}$. Nilpotency of $Q_{sug}$ then gives the
desired KPZ result for the renormalization of the level $k$.

This latter observation can in principle be applied directly to the
case of $W_3$ gravity. The main difficulty now is to find
the spin-three analogue of $T_{ds}$. Here we run into trouble
since (to our knowledge) there is no such spin-three field
that decomposes into a current and a ghost part,
and that forms a closed quantum $W_3$ algebra together with
the $Sl(3,\re)$ generalization of $T_{ds}$.
It was suggested by Feigin and Frenkel \cite{FF} that a better
starting point for the generalization of the above to the case
of $W_3$ is provided by another stress-energy tensor,
namely\footnote{In fact, the stress-energy tensor considered
by Feigin and Frenkel equals $T_{FF}$ without the $kJ_+$ term. This
does not alter the algebra formed by $T_{FF}$, but it should be noted
that without the $kJ_+$ term $T_{FF}$ is {\em not}\, $Q_{ds}$ closed.}
\be \label{better}
T_{FF}=\frac{1}{k+2}\left( :\hat{J}_0\hat{J}_0:+k J_++
(k+1) \dbar \hat{J}_0 \right),
\ee
where we defined $\hat{J}_0=J_0+\beta\gamma$.
In the next section we will show that for the $Sl(3,\re)$
generalization of this stress-energy tensor there is a spin three
field such that they form a quantum $W_3$ algebra.

The following two questions arise: what is the relation between
$T_{FF}$ and $T_{ds}$, and can $T_{FF}$ also be derived from the
quantization of the covariant action? The answer
to the first question is straightforward: $T_{FF}$ differs from
$T_{ds}$ by a $Q_{ds}$ exact term, so it is simply a
different representative of $H_{Q_{ds}}$. In formula we have
\be \label{diffreps}
T_{FF}=T_{ds}+Q_{ds}(\beta J_+).
\ee
The answer to the second question is in the affirmative. To see
this we have to reexamine the ghost redefinitions (\ref{newshoes}).

\newsubsection{Ghost redefinitions}

Recall that we had to introduce new ghost fields in section~2.1
(see (\ref{newshoes})) in order to diagonalize
the ghost action and obtain a recognalizable form for the BRST
charge $Q$. These ghost redefinitions might have seem an
ad hoc trick, whose systematics is unclear.
In this subsection we follow a different procedure
to construct the BRST charge, making it clear
where the ghost redefinitions stem from. Obviously, ghost
redefinitions lead to different expressions for the BRST charge $Q$.
We show that a particular ghost redefinition brings
the BRST charge into the form $Q=Q_{ds}+Q_{FF}$, with
\be \label{qff}
Q_{FF}=\oint :\bgamma\left( T_{mat}+T_{FF}+\hf
T_{gh}^{\bbeta,\bgamma}\right):.
\ee

First, let us review the construction of the BRST charge for
general algebras, linear and non-linear, due to Fradkin and
Fradkina \cite{fradkwad}. Suppose we have a set of constraint
equations $G_a=0$, where the $G_a$ have Poisson
brackets\footnote{For notational simplicity the sub and
superscripts denote generalized indices in the following;
besides labeling the set of constraints they also label the
points in space-time. This suppresses delta-functions and
corresponding integrals in what follows.}
\be \label{struccoeff}
\{ G_a,G_b\}=C^c_{ab}G_c.
\ee
Here the $C^c_{ab}$ are structure {\em functions}, \ie\ they
can be field dependent. If they are field independent, the constraints
form a true algebra, otherwise the algebra is called `soft'.
Introducing for every constraint $G_a$ a ghost $c^a$ and antighost
$b_a$, which satisfy the Poisson brackets
\be \label{spook}
\{ b_a,b_b\}=\{c^a,c^b\}=0,\hspace{1cm}\{ b_a,c^b\}=\del_a^b,
\ee
the classical BRST charge takes the form \cite{henn}
\be \label{general}
Q=\sum_{n\geq 0}(-)^nc^{b_{n+1}}\ldots c^{b_1}
\U{1}{n}{1}{n+1}{U^{(n)}}b_{a_n}\ldots b_{a_1},
\ee
where the $\U{1}{n}{1}{n+1}{U^{(n)}}$ are higher-order
structure {\em functions}, which can be determined from the
following two requirements: $i)$ ${U^{(0)}}_a=G_a$, and $ii)$
$\{ Q,Q \}=0.$ Working out the latter condition, one obtains the
following
recursive formula for the higher order structure functions \cite{henn}:
\ba \label{rec}
(n+1)\,\U{1}{n+1}{1}{n+2}{U^{(n+1)}}G_{a_{n+1}}=\hf
\sum_{p=0}^n(-)^{np+1}\{ \U{1}{p}{1}{p+1}{U^{(p)}},
\U{p+1}{n}{p+2}{n+2}{U^{(n-p)}}\} \nonu
+\sum_{p=0}^{n-1}(-)^{np+n}(p+1)(n-p+1)\,
{U^{(p+1)}}^{a_1\ldots a_pa}_{b_1\ldots b_{p+2}}
{U^{(n-p)}}^{a_{p+1}\ldots a_n}_{b_{p+3}\ldots b_{n+2}a},
\ea
where one should antisymmetrize the r.h.s. in both the $a$ and
$b$ indices. The first order term follows trivially:
${U^{(1)}}^c_{ab}=-\hf C^c_{ab}$. In the case of a true algebra
the $C^a_{bc}$ are constants, from which it follows that
the higher order structure functions vanish.
The BRST charge then takes the familiar form
\be \label{brst5}
Q=c^aG_a-\hf b_cC^c_{ab}c^ac^b.
\ee
For `soft' algebras the higher order structure functions are in general
non-vanishing. Below we will see an example of such an algebra.

Redefining the ghost fields $c^a$ changes the BRST charge as
follows: transforming $c^a\rightarrow \tilde{c}^a=
V^a_bc^b$, with $V^a_b$ some invertible matrix which may depend
on the fields, is equivalent to
defining new generators $\tilde{G}_a=(V^{-1})^b_aG_b$.
(So that the lowest order contribution to the BRST charge
remains the same.)
The new antighost fields $\tilde{b}_a$ follow from
the requirement that they have `diagonal' Poisson
brackets with the $\tilde{c}^a$.
Denoting the structure functions corresponding to the $\tilde{G}_a$
by $\widetilde{U}^{(n)}$, the new BRST charge becomes:
\be \label{newgeneral}
Q=\sum_{n\geq 0}(-)^n\tilde{c}^{b_{n+1}}\ldots \tilde{c}^{b_1}
{\hu^{(n)^{a_1\ldots a_n}}}_{\hspace{4mm}b_1\ldots b_{n+1}}
\tilde{b}_{a_n}\ldots \tilde{b}_{a_1}.
\ee
Note that only the lowest-order contribution to the BRST charge
$c^aG_a=\tilde{c}^a\tilde{G}_a$ is invariant under the ghost
redefinitions.

Let us apply the above considerations to the quantization of
two-dimensional gravity. If we impose the chiral gauge
$\mub=\phi=0$, we have the following constraints:
\ba \label{symms}
\theta = -\frac{\pi}{2}\vars{S_{cov}}{\phi}
& = & J_--k,\nonu
T =  \pi\vars{S_{cov}}{\mub}
& = & T_{mat}+T_{sug}-\mu\theta-z\dbar\theta,
\ea
where we used the equations of motion $\om=0$ and $\bom=-\hf
\dif \mu$, and $T_{sug}$ is given by (\ref{sugimpr}).
To make contact with the previous subsection, we want
(instead of $T$) $T_{mat}+T_{sug}$ as one of the generators.
The BRST charge to lowest order in the
ghost fields reads: $Q=\oint c\theta+\bc T=\oint
(c-\mu\bc+z\dbar\bc)\theta+\bc (T_{mat}+T_{sug})$.
In other words the redefinitions of the
ghost fields $c\rightarrow\gamma=
c-\mu\bc+z\dbar\bc$ and $\bc\rightarrow\bgamma=\bc$, originate
from the corrections in (\ref{symms})
to $T_{mat}+T_{sug}$.
As we explained the corresponding change for the antighost variables
$b\rightarrow\beta$, $\bb\rightarrow\bbeta$ as in (\ref{newshoes})
simply follow by demanding that they have `diagonal' Poisson
brackets with $\gamma,\bgamma$. The complete BRST charge follows
from the algebra formed by $\theta$ and $T_{tot}=T_{mat}+T_{sug}$:
\ba \label{alg}
\{ \theta(\bz),\theta(\bw) \}&=&  0, \nonu
\{ T_{tot}(\bz),\theta(\bw) \}&=&
\theta'(\bw)\del(\bz-\bw),\\[.5mm]
\{ T_{tot}(\bz),T_{tot}(\bw) \}&=&
\deel{1}{12}(d-6k)\del'''(\bz-\bw)+2T_{tot}(\bw)\del'(\bz-\bw)+
T'_{tot}(\bw)\del(\bz-\bw).\nonumber
\ea
Note that in this case we are dealing with a true algebra, \ie\
the $C^a_{bc}$ are field independent, and
using (\ref{brst5}) we can construct the full BRST charge $Q$ as
\be \label{brst3}
Q=\oint \gamma\theta+\bgamma\left( T_{tot}+\hf
T_{gh}^{\bbeta,\bgamma}+T_{gh}^{\beta,\gamma}\right),
\ee
which is the result we found before, see (\ref{brst1}).

The above ghost redefinition occurred because we insisted on
having $T_{mat}+T_{sug}$ as one of the generators. We can
also take a different redefinition such that our BRST charge
becomes $Q=Q_{ds}+Q_{FF}$. For this define yet another set
of ghost fields:
\ba \label{newestshoes}
\zeta=\gamma+J_+\bgamma,&\hspace{1cm}&\bzeta=\bgamma, \nonu
\etab=\bbeta-J_+\beta,&\hspace{1cm}&\eta=\beta.
\ea
This redefinition implies that the new generators are given by
$\theta$ and $T_{new}$, where
\be \label{tnew}
T_{new}=T_{mat}+\frac{1}{k}J_0^2+J_++\dbar J_0.
\ee
The symmetry algebra formed by $T_{new}$ and $\theta$
is slightly different from the previous one, namely
\ba \label{alg2}
\{ \theta(\bz),\theta(\bw) \}&=&  0, \nonu
\{ T_{new}(\bz),\theta(\bw) \}&=&
-\theta(\bw)\del'(\bz-\bw)+\deel{2}{k}
J_0(\bw)\theta(\bw)\del(\bz-\bw),\\[.5mm]
\{ T_{new}(\bz),T_{new}(\bw) \}&=&\deel{1}{12}(d-6k)\del'''(\bz-\bw)+
2T_{new}(\bw)\del'(\bz-\bw)+T'_{new}(\bw)\del(\bz-\bw).\nonumber
\ea
{}From this the full classical BRST charge can be constructed: we find
\be \label{gevonden}
Q=Q_{ds}+Q_{ff}=\oint \zeta\theta+\bzeta\left(
T_{mat}+T_{ff}+\hf T_{gh}^{\etab,\bzeta}\right),
\ee
where $T_{ff}$ is the classical version of $T_{FF}$:
$T_{ff}=\frac{1}{k}\hat{J}_0^2+J_++\dbar \hat{J}_0$,
with $\hat{J}_0=J_0+\eta\zeta$. To extend this result to the quantum
level we should normal order the BRST charge and again allow for
possible renormalizations of $T_{ff}$, \ie\ we  define
\be \label{tochmaar}
T_{FF}=N_T \left( :\hat{J}_0\hat{J}_0:+kJ_++x\dbar\hat{J}_0\right).
\ee
Using the OPE's of the currents and the ghosts, we find that if
we choose
\be \label{cond2}
x=k+1,\hspace{1cm}N_T=\frac{1}{k+2},\hspace{1cm}
d=13+6\left( (k+2)+\frac{1}{(k+2)}\right),
\ee
the quantum BRST charge $Q=Q_{ds}+Q_{FF}$ is nilpotent;
in fact
\be \label{nogsteeds}
Q_{ds}^2=Q_{ds}Q_{FF}+Q_{FF}Q_{ds}=Q_{FF}^2=0.
\ee

\newsection{Quantization of $W_3$ Gravity}

In this section we apply the above machinery to the $W_3$ case.
As before we start with the covariant action and then impose the
chiral gauge. We derive the constraints in this gauge,
and construct (making use of ghost field redefinitions)
the BRST charge corresponding to these constraints.
This BRST charge again
has a natural interpretation in terms of the Drinfeld--Sokolov
reduction for $Sl(3,\re)$. Nilpotency of this charge will lead
to the desired KPZ formula for $W_3$ gravity.

The covariant action for $W_3$ gravity is again given by the
general formula (\ref{pr}), now specialized to the case of
$Sl(3,\re)$:
\be \label{specialidados}
S_{cov}=\gam[\mu,\nu]+\Delta S[\mu,\nu,\mub,\nub,G]+
\gam[\mub,\nub].
\ee
Taking the following Gauss decomposition for $G$
\be \label{chausss}
G=\mat{1}{0}{0}{\om_1}{1}{0}{\om_3}{\om_2}{1}
\mat{e^{\phi_1}}{0}{0}{0}{e^{\phi_2-\phi_1}}{0}{0}{0}{e^{-\phi_2}}
\mat{1}{-\bom_1}{-\bom_3}{0}{1}{-\bom_2}{0}{0}{1}.
\ee
the local counterterm becomes\footnote{In this counterterm
the auxiliary fields $\om_3,\bom_3$ are already integrated out.}
\cite{jj}
\ba \label{lokaalcounteren}
\Delta S = & &\hspace{-5mm}\deel{k}{\pi}\int d^2 z\,\,\Bigl[
       A^{ij}(\om_i+\dif\phi_i)(\bom_j+\dbar\phi_j)
-\sum_i e^{-A^{ij}\phi_j}-\hf A^{ij}\dif\phi_i\dbar\phi_j
\hspace{3cm}\\[.5mm]
  &   & \hspace{13mm}+e^{\phi_1-2\phi_2}(\mu-\hf \dif \nu -\nu\om_1)
      (\mub+\hf \dbar \nub +\nub\bom_1)+e^{-\phi_1-\phi_2}\nu\nub\nonu
  &   & \hspace{13mm}
      +e^{\phi_2-2\phi_1}(\mu+\hf \dif\nu+\nu\om_2)(\mub-\hf \dbar\nub
      -\nub\bom_2) +\mu T+\nu W + \mub \bar{T}
      + \nub \bar {W}\Bigr], \nonumber \label{finalres}
\ea
where $A^{ij}$ is the Cartan matrix of $Sl(3,\re)$, $k$ is
related to the matter central charge $d$ via $d=24k$, and
we defined $T,W,\bar{T},\bar{W}$
through the Fateev-Lyukanov \cite{fatly} construction:
\ba
(\dif-\om_2)(\dif-\om_1+\om_2)(\dif+\om_1) & = &
\dif^3-T\dif-W-\hf \dif T, \nonumber \\
(\dbar-\bom_2)(\dbar-\bom_1+\bom_2)(\dbar+\bom_1) & = & \dbar^3
-\bar{T}\dbar+\bar{W}-\hf \dbar \bar{T}. \label{fat}
\ea
Note that the auxiliary fields $\om_i,\bom_i$ appear in higher than
quadratic form in $S_{cov}$, making it impossible to integrate them
out at this level. $S_{cov}$ is invariant under left and right $W_3$
transformations whose precise form are given in Appendix~B.
The chiral action $\gam[\mu,\nu]$ is related to a $Sl(3,\re)$
WZW model by\footnote{In fact to make contact with the
WZW model, we should define \cite{kj1,jj3}
$\exp(-\gam[\mu,\nu])=\left< \exp(\deel{1}{\pi}\int\mu
T_{mat}+\alpha \nu W_{mat})\right>$, with $\alpha^2=-5/2$.
Of course, a
similar correction should then be taken into account in the
definition of $\gam[\mub,\nub]$.}
\be \label{giraal}
\exp(-\gam[\mu,\nu])=\int Dh\,\del(k\mu-F_{\mu}(J))
\del(k\nu-F_{\nu}(J))\exp(-kS^+_{wzw}(h)),
\ee
with \cite{jj2}
\ba \label{deeffen}
F_{\mu}(J) &=& \hf(J_1+J_2)+\deel{1}{2}z(H_0+H_1)-\deel{1}{4}z^2
(K_1+K_2),\nonu
F_{\nu}(J) &=& J_3+z(J_1-J_2)+\deel{3}{2}z^2(H_0-H_1)
-\hf z^3(K_1-K_2)+\deel{1}{4}z^4K_3,
\ea
where we decomposed the $Sl(3,\re)$ current $J=kh^{-1}\dbar h$ as
\be \label{driedecom}
J=J_aT^a=\mat{H_0}{J_1}{J_3}{K_1}{H_1-H_0}{J_2}{K_3}{K_2}{-H_1}.
\ee

\newsubsection{The chiral gauge}

Let us now impose the chiral gauge $\mub=\nub=\phi_1=\phi_2=0$.
In this gauge the cubic term in the auxiliary fields $\bom_i$
disappears, and the equations of motion for the auxiliary
fields can be solved,
\ba \label{opl}
\om_1=0,&\hspace{1cm}&\bom_1=-\dif\mu-\deel{1}{6}\dif^2\nu,\nonu
\om_2=0,&\hspace{1cm}&\bom_2=-\dif\mu+\deel{1}{6}\dif^2\nu.
\ea
Taking the functional derivative of $S_{cov}$ w.r.t. $\phi_1$ and
$\phi_2$ we obtain the constraints corresponding to the
gauge-fixing of the $\phi$'s:
\ba \label{phigfs}
\pi\vars{S_{cov}}{\phi_1} &=& K_2-2K_1+k-3zK_3=0,\nonu
\pi\vars{S_{cov}}{\phi_2} &=& K_1-2K_2+k+3zK_3=0,
\ea
where we used the above equations of motion
(\ref{opl}) and the relation (\ref{deeffen})
between $\mu,\nu$ and the currents $J_a$.
Since the currents $K_i$ are anti-holomorphic,
the above two constraints lead in fact to the following
{\em three}\, constraints on the currents $K_i$
\be \label{dscon}
K_1=K_2=k\hspace{5mm}\mbox{and}\hspace{5mm}K_3=0.
\ee
So the $Sl(3,\re)$ current $J$ is constrained to
\be \label{fix}
J_{fix}={\cal J}+k\Lambda=
\mat{H_0}{J_1}{J_3}{k}{H_1-H_0}
{J_2}{0}{k}{-H_1},
\ee
where ${\cal J}$ is the matrix containing
the currents $H^i,J^i$ and
$\Lambda$ is the constant matrix with zeroes everywhere except
$\Lambda_{21}=\Lambda_{32}=1$.

The constraints corresponding to the gauge-fixing of $\mub,\nub$ read
\ba \label{ressymm}
T\equiv\pi\vars{S_{cov}}{\mub}&=&T_{mat}+T_{\Delta},\nonu
W\equiv\frac{1}{\sqrt{-\beta_{mat}}}\frac{\pi}{\alpha}
\vars{S_{cov}}{\nub}&=&\pm i \frac{W_{mat}}{\sqrt{\beta_{mat}}}
+\frac{W_{\Delta}}{\sqrt{\beta_{\Delta}}},
\ea
where the factor $\alpha^{-1}=\sqrt{-2/5}$ is explained
in the footnote above. Following \cite{cern} we divided by
an extra factor $\sqrt{-\beta_{mat}}$, which has proven to
be very useful in the construction of the BRST charge for
$W_3$. Since classically we have $\beta_{mat}=16/5c_{mat}$
and $c_{\Delta}=-c_{mat}$, as we shall see shortly,
the matter spin-three
field contains an extra factor $\pm i$ when compared to
$W_{\Delta}$. From now on the rescaled fields $W/\sqrt{\beta}$,
are denoted as $\widetilde{W}$.
Discarding terms proportional to the $\phi_1,\phi_2$ constraints
(\ref{dscon}), since they can be absorbed by making suitable
ghost redefinitions,
we find that $T_{\Delta}$ and $W_{\Delta}$ are given by
\ba \label{tenw}
T_{\Delta} &=& \frac{1}{2k}\tr(J_{fix}^2)+\tr(PJ_{fix})', \nonu
\alpha W_{\Delta} &=& \frac{1}{3k^2}\tr(J_{fix}^3)+\frac{1}{2k}
\tr(PJ_{fix}^2)'+\frac{1}{2} \tr(P^2J_{fix})''\nonu
& &+\frac{1}{4k}\left[ \tr(PJ_{fix})\tr(P^2J_{fix})'-
\tr(P^2J_{fix})\tr(PJ_{fix})'\right],
\ea
where the prime denotes derivation w.r.t. $\bz$, and
we defined $P=\mbox{diag}(1,0,-1)$, satisfying $[P,\ll]=\ll$.
They form a {\em classical}\, $W_3$ algebra with central charge
$c_{\Delta}=-24k=-c_{mat}$. (Note that the factor in front of the
$\frac{1}{3}\tr{J_{fix}^3}$ term equals $\sqrt{3\beta_{\Delta}/k^3}$,
and that the lower order terms pick up a factor $k$ for each
derivative.)

At this point we are in the position to determine the BRST
charge. We have five generators: $G_1=K_1-1$, $G_2=K_2-1$
and $G_3=K_3$ following from the $\phi_1,\phi_2$ gauge-fixing, and
$G_4=T$, $G_5=W$ following from the $\mub,\nub$
gauge-fixing respectively. Using the general scheme of Fradkin and
Fradkina explained in the previous section we can now
compute the BRST charge. The result takes the form
\be \label{vormpje}
Q=Q_{ds}+Q_{W_3},
\ee
where
\ba \label{final}
Q_{ds} &=& \oint c_1(K_1-k)+c_2(K_2-k)+c_3K_3-c_1c_2b_3,\nonu
Q_{W_3} &=& \oint \bc_1\left( T_{mat}+T_{ff}+\hf T_{gh} \right)
+\bc_2\left( \pm i\widetilde{W}_{mat}+\widetilde{W}_{ff}+\hf
\widetilde{W}_{gh}\right).
\ea
This can be established as follows.
The last term in $Q_{ds}$ results from the commutator
between $G_1$ and $G_2$. From the sub-algebra
formed by $T$ and $G_1,G_2,G_3$
\ba \label{res1}
\{ T(\bz),G_1(\bw) \} &=& -G_1\del'(\bz-\bw)
+\deel{1}{k}(2H_0-H_1)G_1\del(\bz-\bw), \nonu
\{ T(\bz),G_2(\bw) \} &=& -G_2\del'(\bz-\bw)
+\deel{1}{k}(2H_1-H_0)G_2\del(\bz-\bw), \nonu
\{ T(\bz),G_3(\bw) \} &=& -2G_3\del'(\bz-\bw)
+(\deel{1}{k}(H_0+H_1)G_3+G_2-G_1)\del(\bz-\bw),
\ea
one deduces the first-order structure functions ${U^{(1)}}^b_{4a}$.
We see that the algebra of the constraints is `soft'.
Using the recursion formula (\ref{rec}) one computes the second-order
structure functions ${U^{(2)}}^{cd}_{4ab}$ as:
\be \label{someuus}
{U^{(2)}}^{12}_{412}={U^{(2)}}^{13}_{413}={U^{(2)}}^{23}_{423}=1/6k,
\ee
and the other vanish. From these first and second-order structure
functions we deduce that $T_{ff}$ appearing in (\ref{final}) is
given by
\be \label{tefef}
T_{ff}=\frac{1}{2k}\tr(\hat{J}_{fix}^2)+\tr(P\hat{J}_{fix})',
\ee
\ie\ it is the simply the formula we found before (\ref{tenw})
for $T_{\Delta}$, with $J_{fix}$ replaced by $\hat{J}_{fix}$,
\be \label{hatfix}
\hat{J}_{fix}=
\mat{\hat{H}_0}{\hat{J}_1}{\hat{J}_3}{k}{\hat{H}_1-\hat{H}_0}
{\hat{J}_2}{0}{k}{-\hat{H}_1},
\ee
with \cite{jj3}
\ba \label{replace}
\hat{H}_0=H_0+b_1c_1+b_3c_3,\hspace{10mm}
\hat{H}_1=H_1+b_2c_2+b_3c_3, \nonu
\hat{J}_1=J_1+b_3c_2,\hspace{12mm}
\hat{J}_2=J_2-b_3c_1,\hspace{12mm}
\hat{J}_3=J_3.
\ea
In a similar way one finds that the spin-three field
$\widetilde{W}_{ff}$ occuring in (\ref{final}) is of the same
form as $W_{\Delta}$,
again with $J_{fix}$ replaced by $\hat{J}_{fix}$.

The only problem left is the determination of $T_{gh}$ and
$\widetilde{W}_{gh}$. Due to the fact that $T$ and $W$
as defined in (\ref{ressymm}) do {\em not}\, form a closed $W_3$
algebra (even though $T_{mat}$ and $W_{mat}$, and $T_{\Delta}$
and $W_{\Delta}$ seperately do), we can not use the scheme of
Fradkin and Fradkina to compute these ghost fields.
Instead we have to make use of the transformation rules for
$\mub,\nub$ (\ref{detail}) that leave the covariant action
invariant. Differentiating these rules w.r.t. $\mub,\nub$
gives us the ghost currents
\ba \label{wtweespook}
T_{gh}&=&\var{\mub}(\bb_1\del_{\bc_1,\bc_2}\mub)+\var{\mub}
(\bb_2\del_{\bc_1,\bc_2}\nub) \nonu
\widetilde{W}_{gh}&=&\kappa\var{\nub}(\bb_1\del_{\bc_1,\bc_2}\mub)+
\var{\nub}(\bb_2\del_{\bc_1,\bc_2}\nub),
\ea
where $\del_{\bc_1,\bc_2}$ are the transformations (\ref{detail})
with $\beps$
replaced by $\bc_1$ and $\lamb$ by $\bc_2$, and we inserted
a factor
\be \label{kappa}
\kappa=-\frac{1}{\alpha^2\beta_{mat}}=\frac{1}{5}
(\beta_{mat}^{-1}-\beta_{\Delta}^{-1}),
\ee
to account for the fact that we divided the spin-three generator by
a factor $\alpha\sqrt{-\beta_{mat}}$. This yields the following
expressions for $T_{gh}$ and $\widetilde{W}_{gh}$
\ba \label{wdriespook}
T_{gh}&=&-\dbar \bb_1\bc_1-2\bb_1\dbar
\bc_1-2\dbar \bb_2\bc_2-3 \bb_2\dbar \bc_2,\\[2mm]
\widetilde{W}_{gh}&=&-\dbar \bb_2\bc_1-3 \bb_2\dbar \bc_1
+3\kappa^{-1}\,
\bb_2\bc_2\,(T_{ff}-T_{mat})+3\bb_1\bc_2\,(W_{ff}-W_{mat})\nonumber
\\[1.5mm]
& &+\hf\bb_1\bc_2\,(3\dif T_{mat}-\dif T_{ff})+
\hf\dbar\bb_1\bc_2\,(3T_{ff}-T_{mat})+
2\bb_1\dbar \bc_2\,(T_{mat}-T_{ff})\nonumber \\[1.5mm]
& &+\kappa\left(
\deel{5}{6}\bb_1\dbar^3\bc_2+\deel{5}{4}\dbar\bb_1\dbar^2\bc_2+
\deel{3}{4}\dbar^2\bb_1\dbar\bc_2+
\deel{1}{6}\dbar^3\bb_1\bc_2\right).
\nonumber
\ea
The reason for rewriting $\kappa$ as the difference of
$1/\beta_{mat}$ and $1/\beta_{\Delta}$ is that then
$\widetilde{W}_{gh}$ can be easily generalized to the quantum
case. One should simply the classical expression for
$\beta=16/5c$ by its quantum analogue $\beta=16/(5c+22)$.
The spin three ghost current is different from the one considered in
\cite{cern,cern2}. However, this difference will disappear
after multiplying the ghost current with $\bc_2$, so it gives
the same contribution to the BRST charge.
Having determined all the ingredients that go into the BRST
charge, one can check the following properties of $Q$ using the
Poisson brackets of the currents and ghosts
\be \label{eigensc}
Q^2_{ds}=Q_{ds}Q_{W_3}+Q_{W_3}Q_{ds}=Q_{W_3}^2=0,
\ee
provided that (classically) $c_{mat}+c_{ff}=0$, or equivalently
$d=24k$.

To extend these results to the quantum level we should again
normal-order the BRST charge and redefine $T_{ff}$ and $W_{ff}$ to
\ba \label{redef}
T_{FF} &=& N_T\Bigl(\frac{1}{2}\tr(\hat{J}^2_{fix})+
x\,\tr(PJ_{fix})'\Bigr), \nonu
W_{FF} &=& N_W\Bigl( \frac{1}{3}
\tr(\hat{J}^3_{fix})+\frac{x}{2}
\tr(P\hat{J}^2_{fix})'+\frac{x^2}{2}
\tr(P^2\hat{J}_{fix})''\nonu
& &\hspace{9mm}+\frac{x}{4}\left[
\tr(P\hat{J}_{fix})\tr(P^2\hat{J}_{fix})'-
\tr(P^2\hat{J}_{fix})\tr(P\hat{J}_{fix})'\right]\Bigl).
\ea
where we have to normal order the r.h.s..
Note that we scale every derivative with a factor $x$. Using the
OPE's of the currents and ghosts, one verifies that $T_{FF}$ and
$W_{FF}$ form a quantum $W_3$ algebra with central charge
\be \label{ccc}
c_{FF}=\frac{8k}{k+3}-24k-30=50-24\left(
(k+3)+\frac{1}{(k+3)}\right),
\ee
if we take
\be \label{teek}
x=k+2,\hspace{10mm}N_T=\frac{1}{k+3},\hspace{10mm}
N_W=\sqrt{\frac{3\beta_{FF}}{(k+3)^3}},
\ee
where $\beta_{FF}=16/(5c_{FF}+22)$.

The BRST charge thus reads $Q=Q_{ds}+Q_{FF}$, with
\be \label{final3ff}
Q_{FF}=\oint :\bc_1 \left( T_{mat}+T_{FF}+\hf T_{gh} \right)+
\bc_2 \left( \pm i\widetilde{W}_{mat}+\widetilde{W}_{FF}+
\hf \widetilde{W}_{gh} \right):,
\ee
where we changed the classical value for
$\beta=16/5c$ to its quantum value
$16/(5c+22)$ in the definition of $\widetilde{W}_{mat}$ and
$\widetilde{W}_{gh}$. We find that
\be \label{twocompl}
Q^2_{ds}=Q_{ds}Q_{FF}+Q_{FF}Q_{ds}=Q_{FF}^2=0
\ee
still holds provided we have $c_{mat}+c_{FF}=100$, or
\be \label{wdriekpz}
d=50+24\left( (k+3)+\frac{1}{(k+3)}\right),
\ee
which is the desired KPZ result for $W_3$.

\newsection{The Conformal Gauge}

In this section we compute the BRST charge
in the conformal gauge $\mu=\mub=\nu=\nub=0$.
In this gauge the covariant action reduces to the Toda action
for $Sl(3,\re)$
\be \label{toda}
S_{toda}=-\deel{k}{\pi}\int d^2 z\,\,\Bigl[
\hf A^{ij}\dif\phi_i\dbar\phi_j+\sum_i e^{-A^{ij}\phi_j} \Bigr],
\ee
when we substitute the equations of motion for $\om_i,\bom_i$,
which in this gauge read $\om_i=-\dif \phi_i$ and
$\bom_i=-\dbar \phi_i$.
The constraints following from the gauge-fixing of $\mu,\nu$ are
\ba \label{confcon}
T\equiv\pi\vars{S_{cov}}{\mu}&=&T_{mat}+T_{toda},\nonu
W\equiv\frac{1}{\sqrt{-\beta_{mat}}}\frac{\pi}{\alpha}
\vars{S_{cov}}{\nu}&=&\pm i \frac{W_{mat}}{\sqrt{\beta_{mat}}}
+\frac{W_{toda}}{\sqrt{\beta_{toda}}}.
\ea
Similar constraints follow of course from the gauge-fixing of
$\mub,\nub$. We will consider only
one chiral sector, since the other sector
gives identical results. So the total BRST charge will be given
by $Q_{total}=Q+\bar{Q}$, where $\bar{Q}$ is the
conjugate of $Q$. They trivially anti-commute with each-other.
Using the equations of
motion for the $\om_i,\bom_i$ we find
\ba \label{todatenw}
T_{toda} &=& \frac{1}{2} A_{ij}\dif\phi_i\dif\phi_j+
\sqrt{k} (\dif^2\phi_1+\dif^2\phi_2),\nonu
\alpha W_{toda} &=& \frac{1}{\sqrt{k}}\dif\phi_1(\dif\phi_2)^2+
\hf\dif\phi_1\dif^2\phi_2-\dif\phi_1\dif^2\phi_1-
\hf\sqrt{k}\dif^3\phi_1-(1\leftrightarrow 2),
\ea
where we rescaled the fields
$\phi_i\rightarrow\phi_i/\sqrt{k}$.
The Poisson brackets for the $\phi_i$ following from
(\ref{toda}) are
\be \label{confpbs}
\{ \dif\phi_i(z),\dif\phi_j(w) \}=A^{-1}_{ij}\del'(z-w).
\ee
Using these Poisson brackets one easily verifies that $\{
T_{toda},W_{toda}\}$ form a {\em classical}\, $W_3$ algebra
with central charge $c_{toda}=-24k$. The BRST charge is given by
\be \label{todabrst}
Q_{toda}=\oint \bc_1\left( T_{mat}+T_{toda}+\hf T_{gh} \right)
+\bc_2\left( \pm i\widetilde{W}_{mat}+\widetilde{W}_{toda}+\hf
\widetilde{W}_{gh}\right),
\ee
where the ghost currents are
still given by (\ref{wdriespook}), with $T_{ff},W_{ff}$ replaced
by $T_{toda},W_{toda}$.
Nilpotency of this classical charge holds due to $d=24k$.
This BRST charge is precisely the same as the one recently
constructed in \cite{cern} (see also \cite{cern2}).

The above results are generalized to the quantum level following
the approach stated in the introduction. We should normal-order the
BRST charge and allow for multiplicative renormalizations of
$T_{toda}$ and $W_{toda}$. So we take
\ba \label{quant}
T_{Toda}&=&N_T \left( \frac{1}{2} A_{ij}\dif\phi_i\dif\phi_j+
x\,(\dif^2\phi_1+\dif^2\phi_2)\right),\nonu
W_{Toda} &=&N_W \left(\dif\phi_1((\dif\phi_2)^2+\frac{x}{2}\dif^2
\phi_2-x\dif^2\phi_1)-\frac{x^2}{2}\dif^3\phi_1-
(1\leftrightarrow 2)\right),
\ea
where the r.h.s. should be normal ordered.
These Toda fields form a quantum $W_3$ algebra of central
charge
\be \label{cc}
c_{Toda}=2-24\frac{(k+2)^2}{k+3},
\ee
provided we take
\be \label{condities}
x=\frac{k+2}{\sqrt{k+3}},\hspace{1cm}N_T=1,\hspace{1cm}
N_W=\sqrt{3\beta_{Toda}}
\ee
where $\beta_{Toda}=16/(5c_{Toda}+22)$. The quantum BRST charge
\be \label{finaltoda}
Q_{Toda}=\oint :\bc_1 \left( T_{mat}+T_{Toda}+\hf T_{gh} \right)+
\bc_2 \left( \pm i\widetilde{W}_{mat}+\widetilde{W}_{Toda}+
\hf \widetilde{W}_{gh} \right):
\ee
is nilpotent iff the KPZ result for $W_3$ holds, \ie\
\be \label{laatste}
d+c_{Toda}=100.
\ee
\vspace{2cm}

\noindent {\large\bf Acknowledgement}\\[1cm]
\noindent We thank B. de Wit for carefully reading the
manuscript. This work was financially supported by the Stichting
voor Fundamenteel Onderzoek der Materie (FOM).

\newpage

\startappendix

\newapp

\noindent The WZW actions $S^{\pm}_{wzw}$ are given by:
\be \label{deff}
S^{\pm}_{wzw}(g)=\deel{1}{4\pi}\int_{\Sigma}d^2z\,\tr{g^{-1}\dif
gg^{-1}\dbar g}\pm \deel{1}{12\pi}\int_B \tr{g^{-1}dg}^3.
\ee
They satisfy the following Polyakov--Wiegman identities \cite{powie}:
\ba \label{idd}
S_{wzw}^{+}(gh) &=& S_{wzw}^{+}(g)+S_{wzw}^{+}(h)
+\ac{g^{-1}\dbar g\dif h h^{-1}}, \nonu
S_{wzw}^{-}(gh) &=& S_{wzw}^{-}(g)+S_{wzw}^{-}(h)
+\ac{g^{-1}\dif g \dbar h h^{-1}}.
\ea

\newapp

\noindent The covariant is constructed such that it is invariant under
$W,\bar{W}$ transformations. In this Appendix we give
the $\bar{W}_3$ transformations that leave the
covariant action describing $W_3$ gravity invariant \cite{jj1,jj2}.
One should realize that there are similar transformations of
opposite chirality that also leave the covariant action invariant.
Let $\mbox{ad}_{\Lambda}$ denote the adjoint action of $\Lambda$
on $sl_3(\re)$, \ie\ $\mbox{ad}_{\Lambda}(Y)=[\Lambda,Y]$.
Its kernel $K$ is spanned by $\{ \Lambda,\Lambda^2 \}$, and
$sl_3(\re)$ can be decomposed as $K\oplus \mbox{Im(ad}_{\Lambda})$.
Let $L$ be the linear operator from $sl_3(\re)$ to $sl_3(\re)$,
defined as the inverse of $\mbox{ad}_{\Lambda}$ on
$\mbox{Im(ad}_{\Lambda})$ and extended
to the zero map on the rest of $sl_3(\re)$.
Define the `field-matrices' $F_{\Delta}=T_{\Delta}\bar{\Lambda}+
W_{\Delta}\bar{\Lambda}^2$ and $F_{mat}=T_{mat}\bar{\Lambda}+
W_{mat}\bar{\Lambda}^2$, with $\bar{\Lambda}$ the transpose of
$\Lambda$, and
\ba \label{teesenwees}
T_{mat}=\pi\vars{\gam[\mub,\nub]}{\mub},&\hspace{1cm}&
T_{\Delta}=\pi\vars{\Delta S}{\mub},\nonu
W_{mat}=\frac{\pi}{\alpha}\vars{\gam[\mub,\nub]}{\nub},&\hspace{1cm}&
W_{\Delta}=\frac{\pi}{\alpha}\vars{\Delta S}{\nub}.
\ea
Let $M(\mub,\nub)$ be the matrix which has zero-curvature together
with $F_{mat}$. In \cite{jj1,jj2} it was
shown that $M(\mub,\nub)$ can be expressed in the following way:
\be \label{jjj}
M(\mub,\nub)=\frac{1}{1+L(\dbar+k^{-1}\mbox{ad}_{F_{mat}})}
M_0(\mub,\nub),
\ee
where $M_0(\mub,\nub)=\mub\Lambda+\nub\Lambda^2$, and
$k=d/24$. (Note that $M_0\in K$.)
In fact the curvature of $M(\mub,\nub)$ and $F_{mat}$
is not identically zero, but of the form
\be \label{vvv}
\mat{0}{\gamma_1+\hf\dbar\gamma_2}{\gamma_2}{0}{0}
{\gamma_1-\hf\dbar\gamma_2}{0}{0}{0},
\ee
where the $\gamma$'s are the (chiral) Ward identities for $W_3$, \ie\
\ba \label{watakaids}
\gamma_1 &=& \dif T_{mat}-D_1\mub-
[3W_{mat}\dbar+2(\dbar W_{mat})]\nub,\nonu
\gamma_2 &=& \dif W_{mat}-[3W_{mat}\dbar+(\dbar W_{mat})]\mub-D_2\nub,
\ea
with $D_1$ and $D_2$ the third and fifth order
Gelfand--Dickey operators
\ba \label{gdops}
D_1 &=& -\frac{d}{12}\dbar^3+2T_{mat}\dbar+\dbar T_{mat},\nonu
D_2 &=& -\frac{d}{360}\dbar^5+\deel{1}{3}T_{mat}\dbar^3+
\hf \dbar T_{mat}\dbar^2+\deel{3}{10}\dbar^2 T_{mat}\dbar+
\deel{1}{15}\dbar^3 T_{mat}\nonu
& &-\beta_{mat}(T^2_{mat}\dbar+T_{mat}\dbar T_{mat}).
\ea
Next define $X(\beps,\lamb)$ to be
\be \label{xxx}
X(\beps,\lamb)=\frac{1}{1+L(\dbar+k^{-1}\mbox{ad}_{F_{\Delta}})}
X_0(\beps,\lamb),
\ee
with $X_0(\beps,\lamb)=\beps\Lambda+\lamb\Lambda^2$.
Recall
that the fields $\om_i,\bom_i,\phi_i$ appearing in
$\Delta S$ originated from one $Sl(3,\re)$ valued
field $G$
\be \label{gausss}
G=\mat{1}{0}{0}{\om_1}{1}{0}{\om_3}{\om_2}{1}
\mat{e^{\phi_1}}{0}{0}{0}{e^{\phi_2-\phi_1}}{0}{0}{0}{e^{-\phi_2}}
\mat{1}{-\bom_1}{-\bom_3}{0}{1}{-\bom_2}{0}{0}{1}.
\ee
The transformation of the $\om_i,\bom_i,\phi_i$ and $\mub,\nub$
can now be extracted from the following formula's
\cite{jj1,jj2}:
\ba \label{jajaja}
\del_{\beps,\lamb}G &=& -GX(\beps,\lamb),\nonu
\del_{\beps,\lamb}M(\mub,\nub) &=& \dif
X(\beps,\lamb)+\left[ M(\mub,\nub),X(\beps,\lamb) \right].
\ea
For $\mub,\nub$ these transformulation rules become:
\ba \label{detail}
\del_{\beps,\lamb} \mub &=& \dif \beps-\mub\dbar \beps+\beps\dbar\mub
-\deel{1}{6}\lamb\dbar^3\nub+\deel{1}{4}\dbar
\lamb\dbar^2\nub-\deel{1}{4}\dbar^2\lamb\dbar\nub+\deel{1}{6}\dbar^3
\lamb\nub\nonu
& &+\deel{1}{2k}(\lamb\dbar\nub-\deel{1}{3}\nub\dbar\lamb)T_{\Delta}+
\deel{1}{2k}(\nub\dbar\lamb-\deel{1}{3}\lamb\dbar\nub)T_{mat}\nonu
& &+\deel{1}{3k}\lamb\nub(\dbar T_{\Delta}-\dbar T_{mat})
+\deel{1}{k}\lamb\nub(W_{\Delta}-W_{mat}),\nonu
\del_{\beps,\lamb} \nub &=& \dif \lamb+\beps\dbar\nub-2\nub\dbar\beps
+2\lamb\dbar\mub-\mub\dbar\lamb+
\deel{1}{k}\lamb\nub(T_{\Delta}-T_{mat}).
\ea


\newpage

\begin{thebibliography}{99}
\bibitem{jj} J. de Boer and J. Goeree, \plb{274} (1992) 289.
\bibitem{cern} M. Bershadsky, W. Lerche, D. Nemeschansky and N.
P. Warner, `A BRST Operator for non-critical W-strings',
CERN-TH.6582/92, HUTP-A034/92, USC-92/015, July 1992.
\bibitem{pol1} A. M. Polyakov, \plb{103} (1981) 207.
\bibitem{herman} H. Verlinde, \np{337} (1990) 652.
\bibitem{pol3} A. M. Polyakov, Int. Journ. of Mod. Phys. {\bf
A5} (1990) 833.
\bibitem{jj1} J. de Boer and J. Goeree, \np{381} (1992) 329.
\bibitem{jj2} J. de Boer and J. Goeree, 'The
Covariant Action and Its Moduli Space from Gauge Theory',
THU-92/14.
\bibitem{vannie} K. Schoutens, A. Sevrin and P. van
Nieuwenhuizen, \np{371} (1992) 315.
\bibitem{wreview1} C. M. Hull, 'Classical and Quantum
$W$-Gravity', QMW/PH/92/1.
\bibitem{wreview2} C. N. Pope, 'Lectures on $W$ Algebras and $W$
Gravity', CTP TAMU-103/91, Lectures given at the Trieste Summer
School in High-Energy Physics, August 1991.
\bibitem{wreview3} K. Schoutens, A. Sevrin and P. van Nieuwenhuizen,
'Induced Gauge Theories and $W$ gravity',
ITP-SB-91-54, CERN-TH.6330/91,
LBL-31381, UCB-PTH-91/51, to appear in 'Strings and Symmetries
1991', Stony Brook, May 1991.
\bibitem{kpz} V. G. Knizhnik, A. M. Polyakov and A. B. Zamolodchikov,
Mod. Phys. Lett. {\bf A3} (1988) 819.
\bibitem{matsuo} Y. Matsuo, \plb{227} (1989) 209.
\bibitem{beroog} M. Bershadsky and H. Ooguri, \cmp{126} (1989) 49.
\bibitem{kj1} H. Ooguri, K. Schoutens, A. Sevrin and P. van
Nieuwenhuizen, `The Induced Action of $W_3$ Gravity',
ITP-SB-91/16, RIMS-764 (June 1991).
\bibitem{cern2} E. Bergshoeff, A. Sevrin and X. Shen, `A
Derivation of the BRST Operator for Non-Critical $W$ Strings',
UG-8/92, CERN-TH.6674/92, UCB-PTH-92/27, LBL-32806, September
1992.
\bibitem{DS} V. G. Drinfeld and V. V. Sokolov, J. Sov. Math.
{\bf 30} (1985) 1975, Sov. Math. Doklady {\bf 62} (1981) 403.
\bibitem{FF} B. Feigin and  E. Frenkel \plb{246} (1990) 75;
E. Frenkel, `$W$-Algebras and Langlands-Drinfeld
Correspondence', lecture given at the Carg\`ese Summer School on
`New Symmetry Principles in QFT', July 1991.
\bibitem{horva} Z. Horvath, L. Palla and P. Vecsernyes, Int.
Journ. of Mod. Phys. {\bf A4} (1989) 5261.
\bibitem{itoh} K. Itoh, \np{342} (1990) 449.
\bibitem{kura} T. Kuramoto, \plb{233} (1989) 363.
\bibitem{BS} P. Bouwknegt and K. Schoutens, `${\cal W}$ Symmetry
in Conformal Field Theory', CERN-TH.6583/92, ITP-SB-92-23, to be
published in Physics Reports.
\bibitem{fradkwad} E. S. Fradkin and T. E. Fradkina, \plb{72}
(1978) 343.
\bibitem{henn} M. Henneaux, Phys. Rep. {\bf 126} (1985) 1.
\bibitem{powie} A. M. Polyakov and P. B. Wiegmann, \plb{131}
(1983) 121; \plb{141} (1984) 233.
\bibitem{fatly} V. A. Fateev and S. L. Lukyanov, Int. Journ. of
Mod. Phys. {\bf A3} (1988) 507.
\bibitem{jj3} J. de Boer and J. Goeree, 'The Effective Action of
$W_3$ Gravity to All Orders', THU-92/33.
\end{thebibliography}
\end{document}